\documentclass{aa}
\usepackage{epsfig,graphicx}
\begin{document}


%
   \title{X-ray emission from classical and recurrent
novae observed with ROSAT}

 \author{M. Orio \inst{1} \inst{2} \inst{3}, J. Covington \inst{4},
 H. \"Ogelman \inst{4}}
   \offprints{M. Orio}

   \institute{Osservatorio Astronomico di Torino, Strada Osservatorio,
              20, I-10025 Pino Torinese (TO), Italy
              \and email: orio@cow.physics.wisc.edu
           \and Astronomy Department, 475 N.Charter Str., Madison WI
           53706
              \and
              Physics Department, 1150 University Avenue, Madison WI
              53706, USA}

   \date{Received ; accepted }


\abstract{
 We have analysed  350 pointed and serendipitous observations
of 108 different classical and recurrent novae in outburst and in quiescence,
contained in the ROSAT archive. One aim was to 
search for {\it super-soft X-ray sources} and we found only 3 
of them among post-novae. 
Thus, the super-soft X-ray phase of novae is relatively
 short lived (up to 10 years) and is observed only for up to 20\% of novae.
 Most classical and recurrent novae instead emit {\it hard X-rays}
(in the ROSAT band) in the first months after the outburst, with peak
 X-ray luminosity of a few times 10$^{33}$ erg s$^{-1}$. 
The emission, which we attribute to
shocks in the nova ejecta,  lasts at least $\simeq$ 2 years and 
even much longer under special circumstances (like preexisting circumstellar
material, or a prolonged wind phase).
 We also investigated X-ray emission due to
the  {\it accretion} mechanism in quiescent 
novae.  81 Galactic classical and recurrent novae
were observed at quiescence, and only 11 were detected.
Some of them are variable in X-rays on time scales of years;
the X-ray spectra range from very soft
to hard.  The average X-ray luminosity
is not larger than that of quiescent dwarf novae, even if 
quiescent novae are  at least 10 times more luminous 
at optical wavelengths. There seems to be a {\it missing boundary
 layer problem}: a possible explanation is that boundary layer
radiation in nova systems is  emitted almost entirely in the
  extreme ultraviolet.
 There is no evidence of a large incidence of magnetic systems,
either of enhanced X-ray luminosity in  novae 
observed shortly {\it before} or {\it after} the outburst.
}
\maketitle
%
\section{Introduction}

{\it Classical novae} are cataclysmic
variables (hereafter CV), close binary systems in which
a white dwarf accretes matter from a Roche lobe-filling companion.
 This companion is often a main sequence star (with mass $<$
1 M$_\odot$) and the known orbital periods are mostly 
in the range 2.5--8 hrs. {\it Recurrent
 novae}, so called because of recurrent outbursts every
 10-30 years, are a class  with only 10
known systems. They can be  either CV or symbiotic-like systems,
with a giant secondary and an orbital period of the order of a year.
Both types of {\it novae}, which we will call
hereafter with this comprehensive name, undergo  outbursts of amplitude $\Delta$m=8-15 mag in the
 optical range,  with the recurrent novae outbursts in the lower tail of
the distribution. Novae are the second most energetic phenomenon
in the Galaxy: 
 the total energy emitted is 10$^{44}$-10$^{46}$
 erg.    The outbursts are thought to be  triggered by a thermonuclear
runaway in the hydrogen burning shell at the bottom of the accreted layer.
Even if there is no subsequent shock wave
a radiation driven wind follows, depleting all or part
of the accreted envelope.

X-ray emission in the ROSAT wavelength range is expected because of 
the following mechanisms:
\begin{figure}
\includegraphics[angle=-90,width=8.7cm,clip]{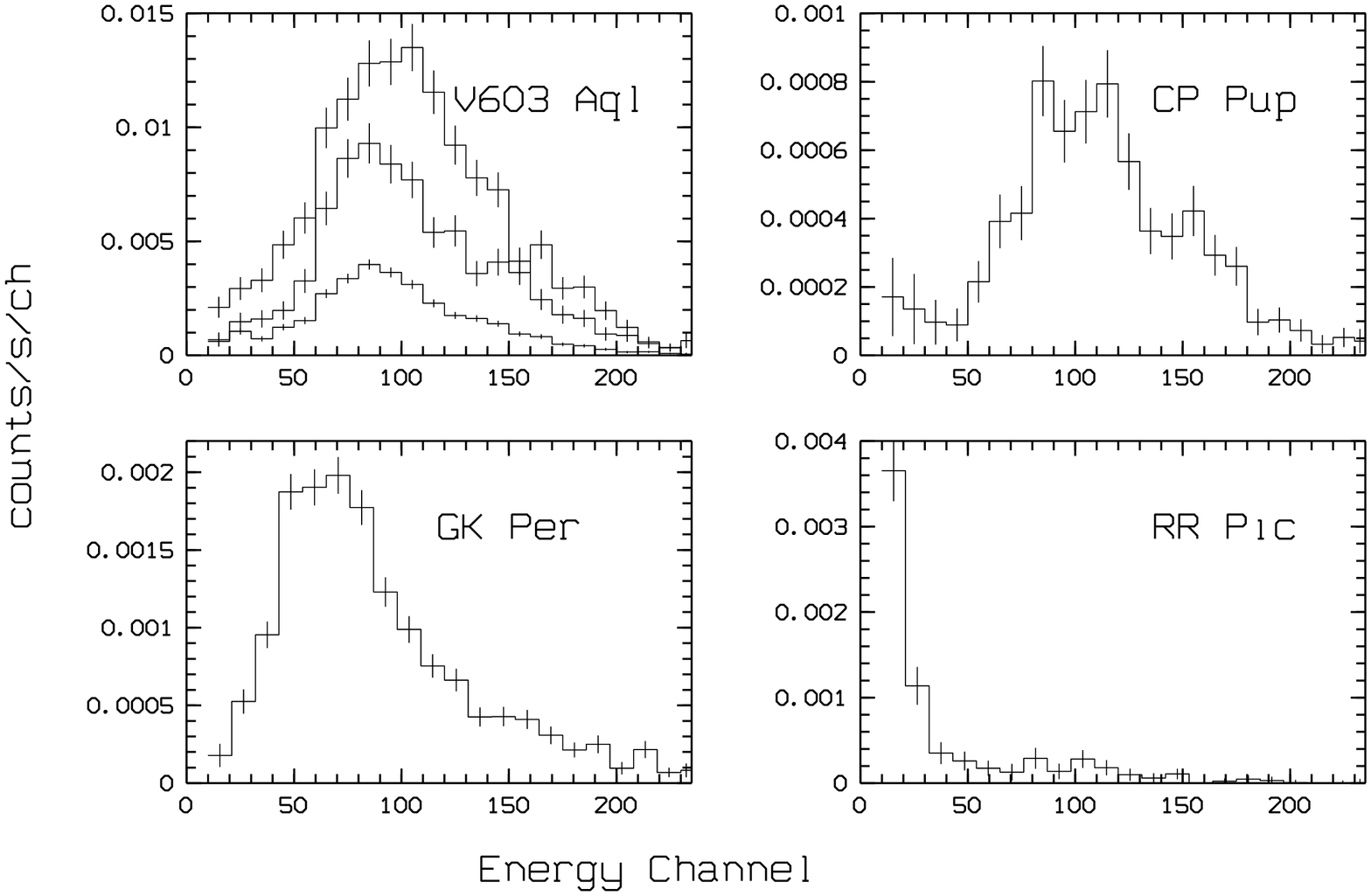}

\medskip
\includegraphics[angle=-90,width=8.7cm,clip]{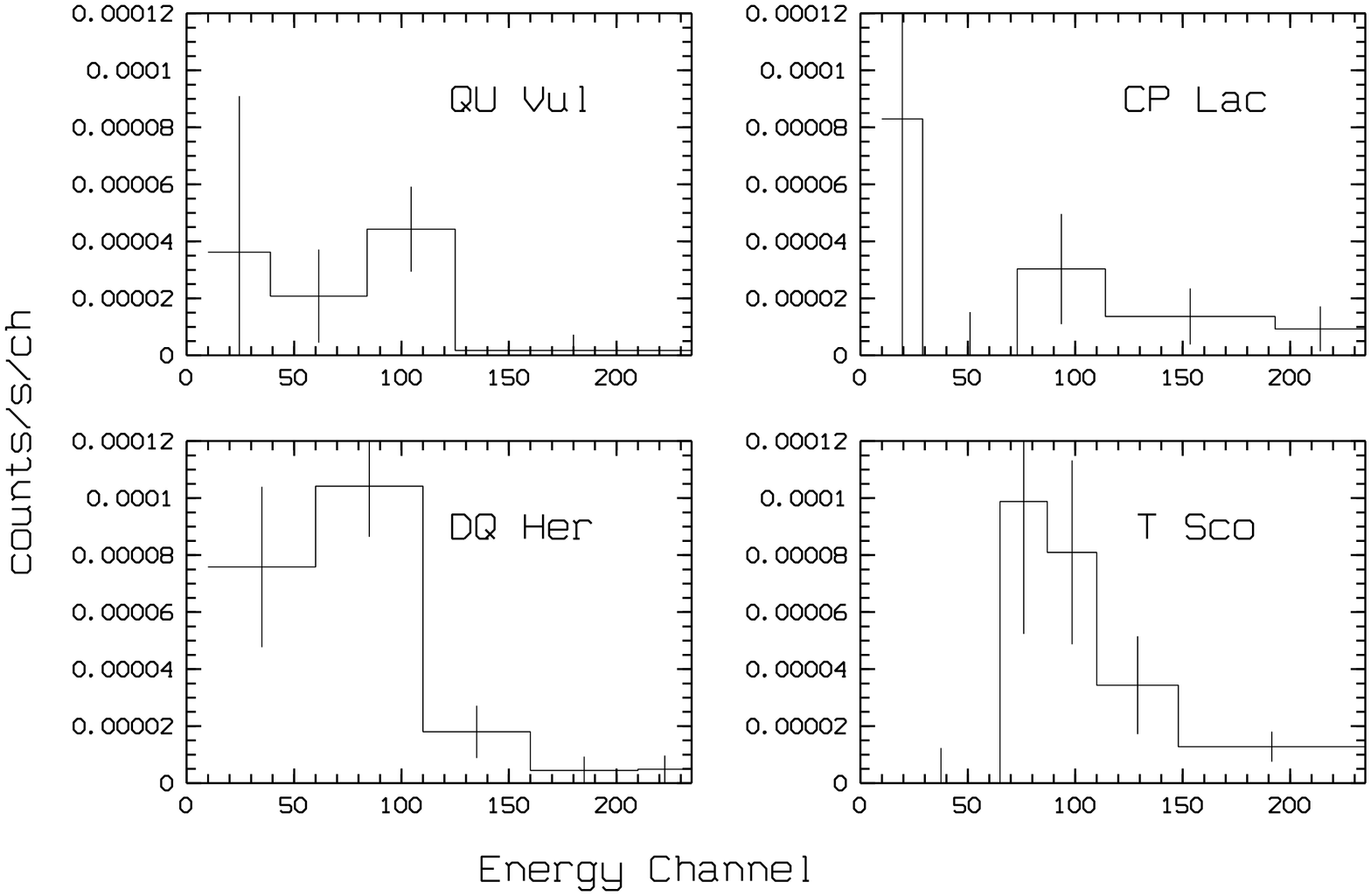}

\medskip
\includegraphics[angle=-90,width=8.7cm,clip]{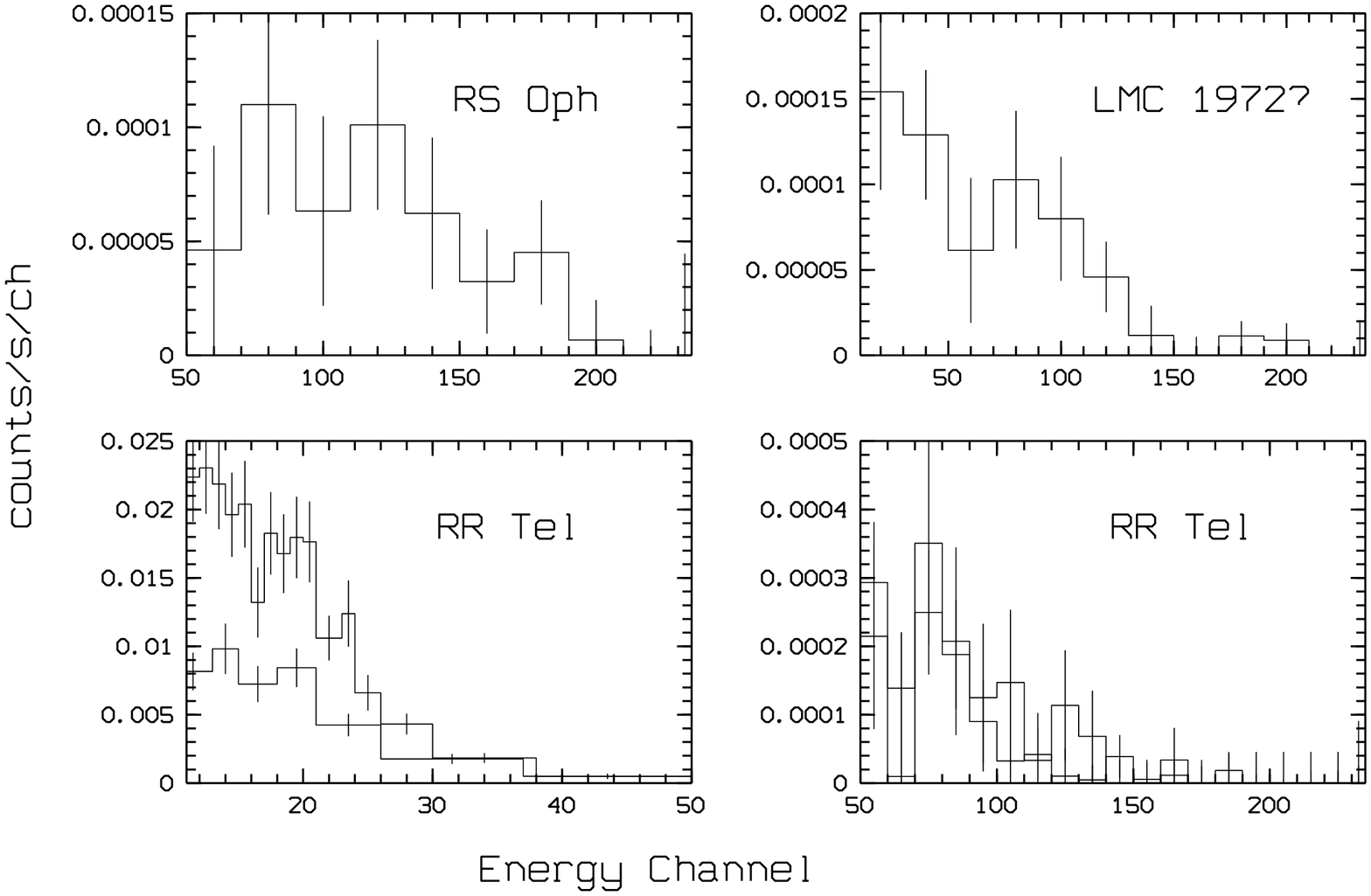}
\caption[spectra]{ROSAT PSPC spectra of 11  quiescent
novae (in two different
observations for V603 Aql and RR Tel). We show the
count rate per energy channel; each of the
channels is approximately equivalent to
10 eV. The  spectra of the brightest
novae in X-rays are on top, except for the one of RR Tel which is
split in two (soft and hard) in the bottom row. The identification of an X-ray
source with N LMC 1972 is uncertain due to large off-axis angles and a crowded
field. }
\label{spectra}
 \end{figure}

1. Shortly after the outburst,
{\it residual hydrogen burning occurs in a shell} on the white dwarf,
  if part of the  accreted envelope is {\it not}
 ejected. The photosphere shrinks
while the effective temperature   increases, hydrogen
burns in a shell and the post-nova 
appears as a very hot black-body-like object  at effective temperatures 2.5-10
$\times 10^5$ K, with L$_{\rm x} \simeq 10^{38}$ erg s$^{-1}$ (the Eddington 
luminosity
for a 1 M$_\odot$ star; see Prialnik 1986). The post-nova at this point is 
a {\it super-soft X-ray source}, and it remains  such for a time
which is 
directly proportional to the leftover envelope mass ( e.g. Kato
1997). The minimum pressure necessary for the
 outburst to be triggered (see Fujimoto 1982) is: 

 $${\rm P}_{\rm outb}={{\rm G} {\rm M}_{\rm WD} \Delta {\rm M}_{\rm env}
 \over 4 \pi {\rm R}^4_{\rm WD}}$$

\noindent  where $\Delta{\rm M}_{\rm env}$ is the mass of the accreted envelope, 
 and M$_{\rm WD}$ and R$_{\rm WD}$ are the mass and radius of the white
 dwarf (WD), respectively.
 The higher the mass of the white dwarf, the earlier  this  pressure
 is reached: less  mass $\Delta$M$_{\rm env}$ is 
accreted on  a massive white dwarf where it is concentrated in
 a smaller surface. Therefore
 $\Delta$M$_{\rm env}\propto{ 1 \over {\rm M}_{\rm WD}}$. Assuming
that  $\Delta$M$_{\rm env}$ is a  fixed fraction of the
accreted envelope, the time to burn its hydrogen content is 
inversely proportional to the original white dwarf mass.

If not all the accreted material is ejected in the outburst,
and if the white dwarf mass  is high (as it is thought to be often the case,
 since outbursts on small mass WD take a much longer time
to trigger and are 
rarer) the  mass increases towards the Chandrasekhar
limit after repeated outbursts. {\it Eventually, the system undergoes
a type I supernova event, or a neutron star is formed by
accretion induced collapse.} The model
predictions are very parameter-dependent and sometimes contradictory
(see Prialnik 1986, Kato 1997).
The large super-soft X-ray flux of the luminous remnant,
  even of a few 10$^{-7}$ erg cm$^{-2}$ s$^{-1}$ for Galactic novae (e.g.
Balman et al. 1998, Orio   1999)  is  the  only
clear  observational evidence of how long the hydrogen
 rich fuel lasts, therefore  of how much is left after each
outburst. Despite the  large interstellar absorption in this
energy range,   
super-soft X-ray source are indeed easily detected in the
Magellanic Clouds, in M31 and sometimes in the Galaxy (see Greiner 2000).

2. During the outburst, the
X-ray flux can be produced by {\it shocks} in
 the hot circumstellar material (Brecher  et al. 1977).
Even if there is no shock wave
in the outburst, shocks may originate in interacting
winds or interaction between the ejecta and
pre-existing material. It has been suggested
 that a high velocity wind  collides
into a lower velocity wind emitted in a previous stage (models are
 developed in Lloyd et al.
1994, 1997, O'Brien et al.  1994, O'Brien \& Lloyd 1994 ). 
 The expected spectrum is
thermal.  The plasma  temperature can be kT=0.2--15 keV,
depending
on how much time has elapsed since the shock, and how efficient the cooling
is. Recurrent novae have a secondary giant star, which
has lost  a significant amount of material
through a wind.  This material is heated and shocked by the nova ejecta.
Hard X-ray emission attributed to shocked gas was first detected   in
 EXOSAT observations of the recurrent nova  RS Oph in its 1985
 outburst (see Mason et al. 1987,  Contini et al. 1995). For classical novae,
shocks are suspected to occur because they
explain the presence of coronal lines in the
nova spectra in the optical (Williams, 1992) and  infrared (Gehrz et al.
1990, Benjamin \& Dinerstein, 1990). 
   \begin{table*}
   \caption[tab1]{Observations of Galactic novae with the ROSAT PSPC:
time elapsed between outburst and observations in years
or in days (d) and  months (m),
E(B-V), distance, time for decay from visual maximum by 3 magnitudes,
ROSAT PSPC count rate $\pm$1$\sigma$ or 3$\sigma$ upper limits,
 image reference number in the HEASARC archive,
off-axis angle  $\alpha$  if the nova was not observed on axis,
reference for d and E(B--V).}
         \label{tab1}
$$
\begin{array}{l c l c l c l c l c l c l c l}
 \hline
 $ Nova$ &   $time  after$  & $E(B-V)$ &  $distance$
  & $t$_3  & $count  rate$  &  $Image$  & \alpha & $Ref.$ \\

         & $outburst$       &            & $(kpc)$ &
 $(days)$ & $(cts s$^{-1})  &  $number$ &  $(arcmin)$  & \\
            \hline
 $1974 Cyg 1992$ & $4-22m$  & 0.18-0.32 & 2-3 & 40 &  $up to $
 76.50 \pm 0.17 & $all$   & (1,2) \\
 &  $63 d$ & &  & & 0.0208 \pm 0.0031 & 141856 & &  \\
 &  $91 d$ & &  & & 0.1426 \pm 0.0110 & 18000 & &  \\
 & $97 d$  & & & & 0.1782 \pm 0.0071  & 18000 & & \\
 & $147 d$ & & & & 0.3516 \pm 0.0122 & 18000 & & \\
 $4157 Sgr 1992a$ & $19 m$ & $0.41-0.51$ & & &  \leq 0.0153 & 400396 & 39 
& (3) \\
 $351 Pup 1991$ & $16 m$ &  0.3 \pm 0.1 & 4.7 \pm 0.6 & 40 &
0.2230 \pm 0.0005 & 300212  & & (4) \\
 $838 Her 1991$ & $6 d$ &  0.53 & 3.4 & 5   & 0.1567 \pm 0.0115
 & 160060  &  & (5)\\
  &   $12 m$   &         &     &   &  \leq 0.0027   & 300162 &  & \\
 & $19 m$  &      &         &   & 0.0009 \pm 0.0003   & 300183 & & \\
 $868 Cen  1991$ & $17 m$  & 1.7 & & &  \leq 0.0035 & 500201 & 29 & \\
 $4160 Sgr 1990$ & $26 m$ & & & &  \leq 0.0046 & 200916 & 37 & (3) \\
 $3890 Sgr  1990 (RN)$ & $11 m$ & 0.5 &  &  10 & 
 \leq 0.0051 & 400168 &  36  & (6) \\
           & $17 m$ & & & &  \leq 0.0013 & 400310 & 22 & \\
 $443 Sct 1989$ & $30 m$ & 0.41  & 8 &  46 \pm 9 &  \leq 0.0042 & 
300124 & 30  & (6,7) \\
& 3  & & & &   \leq 0.0037 & 400316 & 27 &  \\
 $745 Sco 1989 (RN)$ & 2.5  &1.1 & & &  \leq 0.0031 & 900198 & 53 \\
& 2.5 & & & &  \leq 0.0022 & 400151 & 53  & (7) \\
& 3.5 & & & &  \leq 0.0030 & 400269 & 53  & \\
 $394 CrA 1987 (RN)$ & $19.5 m$ & & &  10 &   \leq 0.0011 & 300042 
 &  & (8) \\
$QV Vul 1987$ & 3.25  & 0.32 & 4.5-4.7 &  60 &  \leq 0.0024 & 300040
 &  & (7) \\
            & 4.25  & & & &  \leq 0.0034 & $300040-1$ & \\
 $827 Her 1987$ & $50 m$ & &  2.64 & 57 &    \leq 0.0030 & 300041  & \\
& $62 m$ & & & &  \leq 0.0033 &  $300041-1$ & &  \\
 $OS And  1986$ & 5  &  0.26 \pm 0.04 & 5.1 &  22 &  \leq 0.0008 &
300039  &  & (9) \\
 $Sgr 1986$ & 6  & & & &  \leq 0.0033 & 201093 & 36 &  \\
 $QU Vul 1984$ & 5.3  & 0.57 & 2.6 \pm 0.2 &  34 &
 0.0034 \pm 0.0007 & 300037 & & (10) \\
 $PW Vul 1984$ & 6.9  & 0.55 & 1.6 \pm 0.2 & 147 &  \leq 0.0038 &
300036 &  & (11) \\
& 8.9   & & & &  \leq 0.0020 & $300036-1$ & &  \\
 $4092 Sgr 1984$ & 7.5  & & & & \leq 0.0035 & 400209 & 17 \\
 & 7.5 & & & &  \leq 0.0100 & 400201 & 33  & \\
 & 7.5 & & & &    \leq 0.0113 & 400202 & 45 &  \\
 & 7.5 & & & &  \leq 0.0130 & 400210 & 36 & \\
 & 7.5 & & & &  \leq 0.0119 & $400211-1$ & 55 &  \\
 & 7.5 & & & &  \leq 0.0249 & 400211 & 55  & \\
 $GQ Mus 1983$  & 9  & 0.45 \pm 0.15 & 48 \pm 1 &  45 &
 0.1137 \pm 0.0046 & 300035 & & (12) \\
& 10   & & & & 0.0066 \pm 0.0014 & 300336  & &  \\
& 10.8   & & & &  \leq 0.0015 & 300336 &  & \\
& 11.6   & & & &  \leq 0.0019  & $300336-1$ &  & \\
 $MU Ser 1983$ &10  & & & 5 & \leq 0.0012  & 900170 & \\
 $RS Oph  1985 RN$ & 6 & 0.70 & $1.2-2$ & & 0.0035 \pm 0.0010 & 300038 &
 7 & (6) \\
                & 7 & & & & 0.0116 \pm 0.0017 & $300038-1$ &  & \\
 $4121 Sgr 1983$ & 10 & & & &  \leq 0.0017 & 201093 & 41 & \\
 $SS LMi 1980$ & 12 & & & &  \leq 0.0098 & 700433 & 40  & \\
 $4065 Sgr 1980$ & 14 & & & &  \leq 0.0093 & 500009  & 50 &  \\
 $HS Sge 1977$ & 15 & & &  20 &  \leq 0.0057  & 500190 & 50 &  \\
 & 15 & & & &  \leq 0.0079 & 500191 & 48 & \\
 $2104 Oph 1976$ & 26 & & & &  \leq 0.0083 & 800394 & 52 &  \\
 $NQ Vul 1976$ & 25 &0.37 &2.1 &65 & \leq 0.0077  & 170260 & 56 & &   \\
 $1500 Cyg 1975$  & 26 & 0.69 & 1.2 \pm 0.2 & 3.6  &  \leq 0.0038  &
300033 &  & \\
                & 27 &      &           &   &  \leq 0.0042 &  
 $30033-1$ &  &  \\
 & 27 & & & &  \leq 0.0020 & 900194 & 47 \\
          \noalign{\smallskip}
            \hline
         \end{array}
$$
   \end{table*}
\setcounter{table}{0}
\begin{table*} 
\caption{continued} 
$$
 \begin{array}{l c c c c c c c l}
\hline
 $ Nova$ &   $time  after$  & $E(B-V)$ &  $distance$
  & $t$_3  & $count  rate$  &  $Image$  & \alpha & $Ref.$ \\
 
         & $outburst$       &            & $(kpc)$ &
 $(days)$ & $(cts s$^{-1})  &  $number$ &  $(arcmin)$  & \\
            \hline                                      
 $1301 Aql 1975$ & 26 & & & 35 & \leq 0.0033 & 200126 & 26 & \\
$ Car 1972$ & 20 & & & &  \leq 0.0120 & 200851 & 42  & \\
& 20 & & & &  \leq 0.0053 & 200852 & 23 & \\
& 20 & & & &  \leq 0.0230 & 200857 & 37 & \\
& 20 & & & &  \leq 0.0280 & 200866 & 42  & \\
& 20 & & & &  \leq 0.0057 & 200867 & 22 & \\
 $Car 1972$ & 21 & & & &  \leq 0.0185 & $200872-1$ & 36 &  \\
& 21 & & & &  \leq 0.0232 & 200872 & 36 & \\
 $1330 Cyg 1970$ & 21 & & & 18 &  \leq 0.0064 & 400138 & 46 & \\
 $LV Vul 1968$ & 24 & 0.40 &2.1 & 37 &  \leq 0.0079 & 500086 & 33 & \\
 & 24 & & & &  \leq 0.0097 & 500083 & 33 & \\
 & 24 & & & &  \leq 0.0237 & 500087 & 33 & \\
 $655 CrA 1967$ & 24  & & & &  \leq 0.0020 & 400033 & 14 &  \\
 & 25 & & & &  \leq 0.0021 & $400033-1$ & 14 & \\
 $NSV09828 Sgr 1963$ & 29 & & & 12 &  \leq 0.0100 & 400209 & 37 & \\
 & 29 & & & &  \leq 0.0103 & 400210 & 49 & \\
 $592 Her 1962$ & 30 & & & 27 & \leq 0.0032 & 201228 & 16 &  \\
 $AL Com 1961$ & 30  & & & 30 &  \leq 0.0035 & 700346 & 54&  \\
 $366 Sct 1961$ & 30 & & & &  \leq 0.0105 & 500040 & 37 & \\
  & 30 & & & &  \leq 0.0014 & 500049 & 50 & \\
 $446 Her 1960$ & $33$ &0.037 & 1.38 & 16 & 0.0079 \pm 0.0112 & 600282 & 36 &
  \\
            & 33 & & & & 0.0119 \pm 0.0152 & 600282 & 36 &  \\
            & 33 & & & & 0.0080 \pm 0.0113 &  600283  & 35 &  \\
 $972 Oph 1957$ & 35 & & & 176 &  \leq 0.0051 & 900204 & 47 & \\
 $1275 Sgr 1954$ & 36 & & &  > 10  &  \leq 0.0033 & 200710 & 18 &  \\
 $723 Sco 1952$ & 41 & & &  17  &  \leq 0.0012 & 201383 & 49 & \\
 $2415 Sgr 1951$ & 41 & & & &  \leq 0.0101 & 400210 & 49 & \\
 & 41 & & & &  \leq 0.0074 & 400201 & 25 & \\
 & 41 & & & &  \leq 0.0089 & 400202 & 45 & \\
 & 41 & & & &  \leq 0.0113 & 400187 & 45 & \\
 & 41 & & & &  \leq 0.0150 & 400195 &  31 &  \\
 & 41 & & & &  \leq 0.0123 & 400196 & 48 & \\
 & 41 & & & &  \leq 0.0120 & 400209 & 33 & \\
 & 41 & & & &  \leq 0.0089 & 400202 & 45 & \\
 & 41 & & & &  \leq 0.0092 & 400195 & 31 & \\
 $720 Sco 1950$ & 43 & & & &  \leq 0.0010 & 201383 & 42 & \\
             & 42 & & & &  \leq 0.0010 & 200983 & 42 & \\
             & 43 & & & &  \leq 0.0008 & $200983-1$ & 42 &  \\
 $902 Sco 1949$ & 44 &  & & 200 &  \leq 0.0227 & $500199-1$ & 43 &   \\
 & 43 & & & &  \leq 0.0051 & 500199 & 43 & \\
 $RR Tel 1948$ & 44 & 0.08 & 2.6 &  > 2000 &  0.1666 \pm
0.0060 & 200581 &  & (13,14)\\ 
 & 45 & & & & 0.2297 \pm 0.0104 & 300185 & 19 & \\ 
 $1431 Sgr 1947$ & 45 & & & &  \leq 0.0017 & 600418 & 18 &  \\
 $1150 Sgr 1946$ & 45 & & &  < 600 &  \leq 0.0042 & 500009 & 21 &  \\
 $1431 Sgr 1945$ & 48 & & & &  \leq 0.0017 & 600418 & 18 & \\
 $500 Aql  1943$ & 49 & & &  42 &  \leq 0.0021 & 200898 & 34 &  \\
 $1148 Sgr 1943$ & 50 & & & &  \leq 0.0104 & 400396 & 31 & \\ 
   $CP Pup 1942$ & 50  & $0.1-0.25$ & 0.7 & 8 & 0.0652 \pm 0.0028 &
300212 &  & (15) \\
 $450 Cyg 1942$ & 49 & & & 108 &  \leq0.0036	& 400138 & 52 & \\
 $BT Mon 1939$ & 54 &	 &5.3 & 36? &  \leq 0.0095 & $900266-1$ & 32 &  \\
 & 54 & & & &  \leq 0.0080 & 900266 & 32 & \\
 $CP Lac 1936$ & 57 & 0.48 & 1.15 & 10 & 0.0037 \pm 0.0011 & 20128 & 6 & \\
 $DQ Her 1934$ & 58  &0.07 &0.33 & 94  & 0.0121 \pm  0.0011 & 300103 & & \\
 $737 Sgr 1933$ & 60 & & &  > 70 &  \leq 0.0047 & 201093 & 31 & \\
 $1014 Sgr 1933$ & 60 & & &  > 50 &  \leq 0.0023 & 201093 & 52 & \\
            \hline
         \end{array}
$$
   \end{table*}
\setcounter{table}{0}
\begin{table*}
\caption{continued} 
 \[
 \begin{array}{l c c c c c c c l}
            \hline
 $ Nova$ &   $time  after$  & $E(B-V)$ &  $distance$
  & $t$_3  & $count  rate$  &  $Image$  & \alpha & $Ref.$ \\
 
         & $outburst$       &            & $(kpc)$ &
 $(days)$ & $(cts s$^{-1})  &  $number$ &  $(arcmin)$  & \\
            \hline                                      
 $441 Sgr 1930$ & 61 & & & 53 &  \leq 0.0055 & 300050 & 51 & \\
 $HV Vir 1929$ & 64 & & & &  \leq 0.0028 & 300322 & & \\   
 $1583 Sgr 1928$ & 63 & & & 37  &   \leq 0.0052 & 50009 & 48 & \\ 
 $FM Sgr 1926$ & 65 & & & 30  &  \leq 0.0055 & 500009 & 25 & \\
 $KY Sgr 1926$ & 66 & & & 60 &  \leq 0.0027 & 400330 & 40  & \\
 $RR Pic 1925$ & 68 & $0.02-0.07$  & 0.263 & 150  & 0.0706 \pm 0.0053
 & 300288 & &  \\
 $FL Sgr 1924$ & 47 & & & 32 & \leq 0.0022  & 400151 & 53 & \\
            & 49    & & &    &\leq 0.0029 & 900198 & 53 & \\
 $GR Sgr 1924$ & 67 & & & &  \leq 0.0058 & 300050 & 49 & \\
 $IM Nor 1920$ & 73 & & & &  \leq 0.0010 & 400375 & 24 & \\
 $603 Aql 1918$ & 73 & 0.07 &  0.43 & 8 & 0.3151 \pm 0.0063 & 300056  & & \\
              & 74 &     &       &   &  0.7791 \pm 0.0311 & 300235  & & \\
              & 74 &     &       &   &  1.3035 \pm 0.0321 & 300247  &  &\\
              & 74 &      &      &   &  0.7790-1.3230 & 300255 $ to $ 300266 &
 & \\
 $SS Sge 1916$ & 76 & & & &  \leq 0.0028 & 201100 & 41 & \\
 & 77 & & & &  \leq 0.0247 & $201100-1$ & 41 & \\
 & 76 & & & &  \leq 0.0085 & 500190 & 39  & \\
 & 76 & & & &  \leq 0.0095 & $500190-500191$ & 38 & \\
 $GR Ori 1916$ & 75 & & & &  \leq 0.0024 & 200040 & 40 & \\      
 $GL Sct 1915$ & 77 & & & &  \leq 0.0017 & 200183 & 51 & \\
 $711 Sco 1906$ & 87  & & & & 0.0020 & $200983-1$ & 28 &  \\
 & 87 & & & &  \leq 0.0018 & 201383 & 34  & \\
 $382 Sco 1901$ & 92 & & & &  \leq 0.0009 & 200983 & 44  & \\
 & 92 & & & & \leq 0.0010 & $200983-1$ & 44 & \\
 & 92 & & & &  \leq 0.0010  & 201383 & 36 & \\
 $GK Per 1901$ & 92 & $0.07-0.29$ & 0.575  & 13 & 0.2131 \pm 0.0038 &
 300217 & 41 & \\
 $AT Sgr 1900$ & 93 & & & &  \leq 0.0135 & 400369 & 50 & \\
 $384 Sco 1893$ & 99 & & & &  \leq 0.0065 & 200710 & 35 &  \\
 $T Sco 1860$ & 20 & 0.35 & & 21 & 0.0040 \pm 0.0008 & 300113 & & \\  
 $T Boo 1860$ & 130 & & & &  \leq 0.0010 & 150018 & 24 & \\
 $WY Sge 1793$ & 199 &0.6 & & &  < 0.0015 & 500209 & 46 & \\
 $CK Vul 1670$ & 332 &0.73 & 2.1 & &  \leq 0.0041 & 50086 & 25 &  \\
 & 332 & & & &  \leq 0.0050 & 50083 & 25 & \\
 & 332 & & & &  \leq 0.0082 & 50087 & 25 &  \\     
            \hline
         \end{array}
      \]
\begin{list}{}{}
\item[$^{\mathrm{a}}$]  References: (1) Mathis et al. 1995 and
references therein , (2) Krautter et al.  1996 and references therein,\\
(3) Williams 1994, (4) Orio et al. 1996, (5) Lynch et al. 1992,
(6) Anupama \& Mikolawjeska 1999,\\
  (7) Anupama et al. 1992,
(8) Sekiguchi et al. 1989, (9) Schwarz et al. 1997, (10) Della Valle et
al.
1997,\\ (11) Saizar et al. 1992, and Ringwald
\& Naylor, 1996 (12) \"Ogelman et al. 1991 and
 references therein,\\ (13) Jordan et al. 1994, (14) Whitelock 1988,
(15) MacLaughlin 1960.
{\it Other estimates of distances and absorption  are \\ taken from Harrison \& Gehrz, 1988
and references therein; we caution that
distance estimates given in this work\\ based on the value of t$_3$
are highly uncertain. Values of t$_3$ are from D\"urbeck, 1988, unless
other reference is given.} 
 
\end{list}
   \end{table*}

3. When novae return to {\it quiescence}, {\it accretion}
phenomena produce X-rays. 
For the nova theory, it is critical whether accretion is rekindled,
shortly after the outburst, at a high rate due to the
irradiation of the hot WD, and whether it
is enhanced again before the outburst. This has been predicted by
the ``hibernation theory'' for classical novae (Shara et al. 1986).
The X-ray spectrum of non-magnetic, non-nova CV
is fitted with a  model at   plasma temperatures  kT$\simeq$ 2 keV
(e.g. Richman 1996). Harder components may also exist.
In the standard theory of disk accretion, the boundary layer
is expected to re-radiate
a significant faction of the gravitational energy. 
 How large is this fraction? For other non magnetic CV, Vrtilek et al. (1994)
 estimate it to be about 1/4, and Richman's (1996) 
results are in agreement. However, 
van Teeseling et al. (1996) and Verbunt et al. (1997) argued 
instead that the fraction of re-radiated energy would be so
low that the 
boundary layer model fails to reproduce the X-ray observations. 
According to these authors the X-ray luminosity
is also strongly correlated with the inclination,  
indicating a small size emitting region, closer to the white dwarf
 than the boundary layer.
Observing quiescent novae in X-rays is more
 difficult: CV are usually
discovered in outburst and novae are at least a factor of
100 more  luminous than dwarf novae, so a selection effect exists
in the average distance.
 
 We search also for possible magnetic systems among novae,
because a high  magnetic
  field may have important consequences in the mass ejection process
during outburst (see Orio et al. 1992, Lepine et al. 1999). If a nova
WD accretes
only through the  polar caps (i.e. the system is a  polar) it is
 expected to show
high plasma temperatures, even 20-30 keV (see Cropper 1990,
Patterson 1994). For polar systems, there is  a very soft       
black-body like component.
Often only this component is detected (e.g. Ramsay et al.
1995). Even intermediate polars (IP), however,
may exhibit a very soft component (e.g. Burwitz \&
Reinsch 2000).
The X-ray luminosity of non-magnetic systems is on
average  at least an order of magnitude less than  that of magnetic
systems (10$^{30}-10^{31}$ erg s$^{-1}$  compared to 10$^{32}-10^{33}$
erg s$^{-1}$, e.g. Verbunt et al. 1997).
\begin{table}
   \caption[tabLMC]{Observations of LMC novae with the ROSAT PSPC
and HRI: PSPC count rates or 3$\sigma$ PSPC upper limits, or 
HRI  equivalent count rate (HRI eq). 
Conversion from HRI to PSPC count rates is done according to
Greiner et al., 1996. The LMC 1995 count rate entered is
the highest of all observations (Orio \& Greiner 1999).
We indicate also the number years elapsed between outburst and observations,
number of exposures, the range of off-axis angles $\alpha$ in arcmin. }
         \label{tabLMC}
 \[
 \begin{array}{l c c c l}
\hline
 $ Nova$ & $count rate $ & $ Years$ & $ number $ & \alpha \\
 & $(cts ~ s$^{-1})& $after$ & $of times$ &  \\
                \hline                      
 $LMC  1926$  &   \leq 0.0034 & 64-67 & 3  & 33-40 \\
 $LMC 1936$   &   \leq 0.0024 & 54--57 & 3 & 40-46 \\
 $LMC  1948$  &  \leq 0.0023 & 42-50 & 7 & 20-45 \\
 $LMC  1951$ &  \leq 0.0011 &  39-42 &  1 & 9 \\
 $LMC 1968$ &  \leq 0.0016  & 25 & 1 & 52 \\
 $LMC  1970$ &  \leq 0.0024 &  20-28 & 6 & 13-40 \\
 $LMC  1970b$ &  \leq 0.00 &     21-23 & 4 & 36-43 \\
 $LMC 1971a$ &  \leq 0.0090 & 22-23 & 16 & 36-43 \\
 $LMC 1972?$   & 0.0055 \pm 0.0011 & 18-21 & 30 & 15-46 \\
 $LMC 1973$ &   \leq 0.0017 &  17-20 & 8 & 9-14 \\
 $LMC  1977$  &  \leq 0.0097 &   14-16 & 5 & 47 \\
 $LMC 1978a$ &  \leq 0.0025 &  14-15 & 2 & 48 \\
 $LMC 1978b$ &  \leq 0.0034 &  14-15 & 1 & 37-45 \\
 $LMC  1981$ &  \leq 0.0023 &   9-17  & 6 & 10-27 \\
 $LMC  1987$ &   \leq 0.0040 &  3-7 & 30 & 0-22 \\
 $LMC 1988a$ &  \leq 0.0106 &   2-10 & 20 & 6-37 \\
 $LMC 1988b$ &  \leq 0.0051 &   2-3.8 & 2 & 0-8 \\
 $LMC 1990a$ &  \leq 0.00 & 0.3-9 & 19 & 12-47 \\
 $LMC 1990b$ &  \leq 0.0015 &  3 & 1  & 9-33 \\
 $LMC 1992$ & $(HRI eq) $ 0.0080 & 0.25  & 1 &16-52 \\
                &  \leq 0.0050      & 0.1-10 & 6 & \\
$LMC 1995$  & 0.0610 \pm 0.0003  &  0.3-4 & 8 & 0-44 \\           
 $SMC 1927$ &  \leq 0.0025 & 65-66 & 2 & 45 \\
 $SMC 1951$  &  \leq 0.0018 &  41-43 & 3 & 14-43 \\
 $SMC 1952$ &  \leq 0.0015 & 40-41 & 6 & 20-24 \\
 $SMC 1994$ & $(HRI eq)$ \leq 0.0028 & 2 & 16 & 2-40 \\
            \hline
         \end{array}
      \]
   \end{table}            
Only few X-ray observations of
quiescent accreting novae have been published so far.
 {\it Einstein} IPC detections of a handful of quiescent
novae were described by Becker \& Marshall (1981); Drechsel
et al. (1987) discussed V603 Aql.
 We also reported on the  few detections of quiescent novae
in the ROSAT All-Sky Survey (Orio et al., 1992),
 and analysed  ROSAT pointed observations of RS Oph and CP Pup
(Orio 1993, Balman et al. 1995). Other groups have studied
RR Tel (Jordan et al. 1994),
and mentioned the ROSAT detections of four other novae
in papers dedicated to dwarf novae (Vrtilek et al.
1994, Richman 1996, van Teeseling et al. 1996, Verbunt et al.
1997). In this article,
we examine the ROSAT observations of  {\it all} accreting novae
for a comprehensive view.
\section{The observations}
 
  ROSAT  brought important changes to
X-ray astronomy. It performed the first
All-Sky Survey in soft X-rays and gave            
the scientific community access to a database of  more than
10000 deep pointed observations, covering almost
15\% of the sky and containing  a very  large number of serendipitously
observed sources.  The total number of sources in the web catalogs
 amounts to 95331 (see the ROSAT Consortium web
page, http://wave.xray.mpe.mpg.de/rosat).
 More than half of these sources  were not previously known,
and  a non negligible fraction was observed more than once.
Most observations  used in this work
were done with  the Position Sensitive Proportional Counter, or PSPC,
sensitive in the range 0.2-2.4 keV, with a moderate spectral
resolution ($\Delta$E/E = 0.43 (E/0.93$^{-0.5}$)) and a spatial resolution
of about 25 arcsec at 1 keV.    The field of view of the PSPC
was  2 degrees.  A few observations were also done 
with the HRI (High Resolution Imager), an instrument
 without spectral resolution  but with a good spatial
resolution, of about 2 arcsec, and a 38 arcmin (on a side) square field of view.
 The exposures
in the All-Sky Survey were very short and the sensitivity
limited, therefore  we searched 
 all pointed and serendipitous observations of classical novae
in the ROSAT archive 
(partial results were given in \"Ogelman \& Orio 1995,
Orio \& \"Ogelman 1996, 1998 and Orio 1999).
15 novae were targets of pointed observations; 
only two of them while still  in the nebular phase, and four of them 
 were observed only a few years after the outburst.
 Since novae are mainly observed towards the centre of the Galaxy,
 the number of serendipitous  observations was large: 
  we found archival exposures for 68 other Galactic 
novae. Also 20 novae were repeatedly observed in the Magellanic Clouds.
The whole sample of 103 objects includes also 4 recurrent novae.
The vignetting corrected count rates  or the 3$\sigma$ upper limits 
for the Galactic novae observed
with the ROSAT PSPC are given in Table 1. We also list basic 
information
on the nature of the nova, if this is known: E(B-V), the distance,
and the time to decay by 3 mag from visual maximum, t$_3$
(which is correlated with the ejecta velocity, see Della Valle \& Livio, 1995.
These authors  actually show that the time for a decay by 2 mag, 
 is even  more significant. However it is not known 
for a meaningful number of novae). We
include the off-axis angle $\alpha$ from the centre of the PSPC
and the time elapsed between the outburst and the ROSAT
observation. We caution that 
E(B-V), from which the column of neutral hydrogen is inferred (we
follow Diplas \& Savage 1994), was often measured only during
 outburst and therefore it includes the contribution
of the ejected shell in the first months, while it was still
partially optically thick (e.g.  Orio et al. 1996).
 Table 2 indicates the results for novae repeatedly observed in the Magellanic
Clouds.
The data analysis was performed with the MIDAS-EXSAS package (Zimmermann
et al.  1994). We determined the count rates and the upper limits
with the source detection and upper limits algorithm, 
which also corrects for the presence of the support
structure. In the following sections we also make comparisons 
with observations of other X-ray satellites.
We specially refer to the {\it Einstein} 
IPC (Imaging Proportional Counter), which was sensitive 
in the 0.4-4.0 keV range,
had spectral resolution, a 1 arcmin spatial 
resolution, and a field of view of 75 arcmin. Another satellite used
to observe a few novae was the non-imaging
EXOSAT, with the 
low energy (LE1) telescope and the Channel Multiplier Array (CMA), 
sensitive to the 0.03-2.5 keV region without spectral resolution, a 2$^o$
  field of view and 18 arcsec spatial resolution. 
We indicate approximate conversions of count rates 
 with different satellites using the software
PIMMS, available at the NASA-Goddard Space Flight centre website. 
PIMMS accounts for the different wavelength band-passes.
 Of course, this conversion carries
uncertainties, and it is done only when a physically meaningful 
spectral model can be reasonably assumed.
                            
\section {The missing super-soft X-ray source}

	\begin{table*}
	 \caption[GQ Mus]{Journal of X-ray observations of GQ Mus: vignetting
corrected count rates $\pm$ 1$\sigma$ or 3$\sigma$
upper limits, and parameters derived from spectral fits. MDV-CO refers
 to the model of a CO WD atmosphere described in the text (MacDonald \& Vennes,
1991.)}
			\label{GQ Mus}
  \[
	\begin{array}{l c c c c c l}
 \hline
				\noalign{\smallskip}
  $Observation$ & $Instrument$&  $Count rate$
 &$Model$ &  $T$_{\rm WD} & $Luminosity$ & $Reference$\\
  $date$ & & $(cts ~  s$^{-1}) & & 10^3 $ (K)$ &  (10^{37} $erg 
 s$^{-1}) & \\
  \noalign{\smallskip}
	\hline
	\noalign{\smallskip}
 $April 1984$ & $EXOSAT LE1$ & 0.0034 \pm 0.0009 & & & &  $\"Ogelman et
al. 1984$ \\
  $July 1984$ & $EXOSAT LE1$ & 0.0028 \pm 0.0004 & & & & $\"Ogelman et al.
1987 $\\
 $November 1984$ & $EXOSAT LE1$ & 0.0034 \pm 0.0007 & $blackbody$ &
  280 \pm 50 & 5 \pm 5 &	$\"Ogelman et al. 1987$ \\
 & & & & & \\
 $June 1985$ & $EXOSAT LE1$ & 0.0014 \pm 0.0004 & & & & $\"Ogelman et al. 1987$\\
 & & & & & \\
$July-Aug 1990$ &  $ROSAT PSPC$ & 0.1650 \pm 0.0430 &$blackbody$
 & & &	$Orio et al. 1992$   \\
 & & & & & \\
  $February 1992$ & $ROSAT PSPC$ & 0.1137 \pm 0.0046 & $MDV-CO$ &
430 ^{+10}_{-23}  & 0.15 ^{+1.40}_{-0.05} &  $\"Ogelman et al.
1993$ \\
		 &	  &	   & $blackbody$ & 340 \pm 12 &  \geq 2.6 & \\
 & & & & & \\
 $January 1993$ & $ROSAT PSPC$ & 0.0066 \pm 0.0014 & $MDV-CO$ &
 195 ^{+87}_{-20} &   & $Shanley et al. 1995$ \\
			&		&		   & $blackbody$ & 250
\pm 60 & >0.1 & \\
 & & & & & \\
 $Aug-Sept 1993$ & $ROSAT PSPC$ & \leq 0.0011 & $blackbody$ & \leq 200 & &
$Shanley et al. 1995$  \\
 $July 1994$ & $ROSAT PSPC$ &\leq 0.0012 & $blackbody$ &  \leq 200  &
& $this paper $ \\
		 	\noalign{\smallskip}
	     \hline
	  \end{array}
      \]
	 \end{table*}

Only 3 super-soft X-ray sources are detected among the 
132  Galactic and Magellanic Clouds novae observed with ROSAT.
1974 Cyg and GQ Mus are already well known super-soft sources (see
Introduction, Table 1, and references therein). 
In the course of this project we serendipitously discovered 
also N LMC 1995 ( see Orio \& Greiner, 1999).
Altogether, 30 of the novae in the Galaxy and 9 among
the 25 in the Magellanic Clouds 
were observed within 10 years after the outburst.
After the ROSAT mission, two more classical novae and one recurrent nova were
observed with BeppoSAX and Chandra
and found to be super-soft X-ray sources (Kahabka et al.  1999, Orio
et al. 1999).  Even including these three, only
 $\simeq$15\% of observed post-novae were detectable super-soft X-ray sources, 
at least for a significant length of time to be ``caught'' as such.
Even if E(B-V) is not known
for the majority of the novae, the position in the Galaxy and
the magnitudes reached indicate it is  unlikely
that E(B--V)$>$0.45 for most novae.
With this value, GQ Mus was even detected in a short Survey exposure
(Orio et al. 1992a). There is only one detection in
the Magellanic Clouds, yet
12 other, non-nova super-soft X-ray sources, have been  observed in the LMC
and in the SMC (see Kahabka \& van den Heuvel 1997, Greiner 2000).
Hot post-nova white dwarfs, like GQ Mus or V1974 Cyg, are therefore
relatively rare.  

The 3$\sigma$ upper limits on T$_{\rm BB}$ (equivalent blackbody
temperature), for the novae with known E(B-V) and distance are 
T$_{\rm BB}<$20-35 eV for L$_{\rm bol}  \geq$ 0.1 L$_{\rm Edd}$. 
 At the 1$\sigma$ confidence level,
T$_{\rm BB}<$25 eV in all observations. This is 
much lower than the peak temperature of a shell-hydrogen-
burning white dwarf with M$_{\rm WD} \geq$ 0.8 M$_\odot$ calculated in
several papers (e.g. Prialnik 1986, MacDonald and Vennes 1991).
It is unlikely that all observed novae have lower white dwarf mass,
 or that all have been observed already  during turn off,
while the effective temperature was already   decreasing.
 Turn off in the UV range is also
observed to occur within a few years range (Gonzalez-Riestra
et al. 1998).
   \begin{table}
	\caption[tabaccr1]{X-ray observations of
quiescent novae: count rates
or 3$\sigma$ upper limits, and
PSPC hardness ratio h.r. 
(count rate  at 0.20-0.50 keV/ count rate at 0.51-2.40
 keV, if measurable with
 less than 10\% uncertainty).  For the IPC 
the approximate PSPC equivalent count rate is in parenthesis.}
  \label{tabaccr1}
 \[
 \begin{array}{l c c c l}
\hline
 $Nova$ & $Instrument$ & $c.r. $ & $h.r.$ \\
	    \hline
 ${\rm RR Tel}$ & & & \\
 1978 & $Einstein IPC$  & 0.018 \pm 0.005 (0.06) & \\
 1990 & $PSPC (Survey)$	  & 0.312 \pm 0.053 &  \\
 1992 & $PSPC$	  &  0.167  \pm 0.006 &	 \\
 1993 & $PSPC$	  &   0.230 \pm 0.010 &	 >4   \\
	    \hline
 ${\rm QU Vul}$ & & & \\
  1991 & $PSPC$ & 0.003	 \pm 0.001 & 0.32 \\
	    \hline
  ${\rm RR Pic}$  & & & \\
1979 &  $Einstein IPC$ & 0.031  \pm 0.005 (0.10) &	  \\
 1990 &   $PSPC (Survey)$  &	0.144 \pm 0.043 & \\
 1992-1993 &   $PSPC$  &  0.071 \pm 0.005 & 3.10 \\
	  \hline
  ${\rm V603 Aql}$ & & & \\
     1981 & $Einstein IPC$ & 0.71 \pm 0.03 (1.30) &  \\
     1983 & $Einstein IPC$ & 0.279 \pm 0.013 (0.50) & \\
     1983 & $Einstein MPC $ & 3.24 \pm 0.03 & \\
     1984 & $EXOSAT LE $    & 4.3 \pm 0.7     & \\
     1984 & $EXOSAT LE $    &  7.0 \pm 0.8     & \\
     1984 & $EXOSAT ME $    & 0.75 \pm 0.25    & \\    
 1990 & $PSPC (Survey)$ & 0.536	 \pm 0.046 & \\
 1991 & $PSPC$	& 0.315 \pm 0.006 & 0.12 \\
1993 &  $PSPC$ &  0.8 $ to $ 1.3	 & 0.07 $ to $ 0.11 \\
	\hline
  ${\rm CP Pup}$  & & & \\
1979 & $ Einstein IPC$ & 0.060 \pm 0.006 (0.08) & \\
1990 &  $PSPC (Survey)$  & 0.061 \pm 0.021		& \\
1993 &  $PSPC$	  & 0.065 \pm 0.003		& 0.07 \\
	\hline
  ${\rm GK Per}$ & & & \\
1979-1980				& $Einstein IPC$ & 0.178 \pm 0.016 (0.33) &   \\
1990& $PSPC (Survey)$	 & 0.120 \pm 0.016	    &	\\
1992 & $PSPC$	 & 0.213 \pm 0.004	    & 0.21  \\
1996 &  $ROSAT HRI$	 & 0.065 \pm 0.003		& \\
	\hline
  ${\rm V841 Oph}$ & & & \\
1979&  $Einstein IPC$ & 0.019 \pm 0.005 (0.04) &	 \\
1990& $PSPC$	  & 0.028 \pm 0.010		&
\\
	\hline
  $ {\rm CP Lac}$ & & & \\
1979&  $Einstein IPC$ & \leq 0.012 (\leq 0.03) & \\
	1993& $PSPC$	  & 0.037 \pm 0.001 & 0.65 \\
	\hline
 ${\rm T Sco}$ & & & \\
	1992&  $PSPC$ & 0.0004 \pm 0.0001 & \\
	\hline
  ${\rm DQ Her}$ & & & \\
	1979&  $Einstein IPC$ & \leq 0.030 & \\
	1992 & $PSPC$	   & 0.012 \pm 0.001  & 0.56 \\
	\hline
  ${\rm RS Oph}$ & & & \\
	1991	   & $PSPC$ & 0.012 \pm 0.002 & \\
1992& $PSPC$ & 0.003 \pm 0.001 & \\
		\hline
    ${\rm V446 Her}$ & & & \\
1990	& $PSPC (Survey)$ &	 0.036 \pm 0.012 &  \\
1993 & $PSPC$	& \leq 0.013	&   \\
	\hline
		 \end{array}
      \]
	\end{table}
 There is no trace of residual super-soft X-ray emission from
  QU Vul and  PW Vul, observed to be bright in X-rays in 1987
with EXOSAT (\"Ogelman et al. 1987).
V1500 Cyg, for which ``an almost 3$\sigma$ detection''
with the {\it Einstein} IPC was reported by 
Chlebowski \& Kaluzny (1988) is {\it not} detected,
not even in the 1-2.4 keV range (Verbunt
 et al. 1997, mention
a marginal detection in this range in the All-Sky Survey).
  A firm 3$\sigma$ upper limit is T$_{\rm BB}\leq$26 eV
assuming L$_{\rm x} \geq 10^{36}$ erg s$^{-1}$ and
N(H)$\simeq$ 10$^{21}$ cm$^{-2}$
(compatible with E(B-V)=0.69, Wu \& Kester, 1977).

 In Table 5 we summarize the observations of  GQ Mus, the
only long term X-ray light curve, 
 including an unpublished observation of 1994.
We give count rates, effective temperature and bolometric
luminosity derived with two different models,
including a C-O white dwarf atmospheric model in LTE by MacDonald
\& Vennes (1991), not used in the detection papers. 
Atmospheric models yield lower 
values of the luminosity and  higher T$_{\rm WD}$ than the blackbody
assumption. The model 
was  studied for a white dwarf of  1.2 M$_\odot$ emitting at Eddington
luminosity (L$_{\rm bol}$=1.6 $\times$ 10$^{38}$ erg s$^{-1}$), and
we assume d=4.75$\pm$1.55 kpc
(Krautter \& Williams 1989). 
The bolometric luminosity
is expected to vary by up to a factor of 10 between
1.3 and 0.6 M$_\odot$. It is commonly re-scaled with the mass
using the formula
 L$_{\rm bol}$ = A(M$_{\rm WD}$ -B),  where B is a function of the chemical
composition of the envelope, and ranges between 0.2 and 0.6 (e.g. Truran 1979).

 The EXOSAT LE telescope did not possess
spectral resolution, however the count rates of the first three
observations are consistent with the predictions of
 a blackbody model with the parameters reported in the
table.  The spectrum during the 1990 All-Sky Survey observation was
very soft (Orio et al. 1992a), consistently with 
a blackbody temperature in the range 30-50 eV.
The conversion of EXOSAT into ROSAT count rates 
in the super-soft range is uncertain and we do not
attempt to compare the luminosity. 
In 1992 we were observing the ``plateau'' constant
bolometric luminosity, because the turn-off
 decay of the luminosity is expected
to last only about one year (Prialnik, private communication).
 The 3$\sigma$ range of values of the bolometric
luminosity is below  0.1 L$_{\rm Edd}$ for a 1.2 M$_\odot$ WD
(L$_{\rm Edd}$  as given by MacDonald and Vennes, 1991;
 there is an additional uncertainty of a factor of
2 due to  the 33\% distance uncertainty).
Therefore, the white dwarf mass must be significantly
smaller than 1.2  M$_\odot$,
around the lowest end of the range, 0.6  M$_\odot$.
The 3$\sigma$ lower limit T$_{\rm WD} \geq$400000 K
is also  the peak temperature of a WD of 0.6
M$_\odot$.  By 1993, turn-off started because the effective
temperature was too low for a post-nova
which is still burning hydrogen. We conclude that hydrogen burning
lasted for 8 to 9 years. This is exactly
the length of the constant bolometric luminosity
phase for  a WD mass 0.6  M$_\odot$ according to Kato (1997),
so things seem to fit together.

 RR Tel, a transition object between classical and
 symbiotic novae, is a EUV source with a hard X-ray tail,
and not a super-soft X-ray source in the conventional sense.
  The super-soft   portion of the spectrum varies 
inversely with  the hard X-rays flux (Fig. 1). 
Before repeated ROSAT observations, Jordan et al. (1994) had speculated
instead that only the hard X-ray component varies. We attribute
the large {\it increase} of the super-soft component, 
(while the counts above 0.5 keV {\it decreased})  
 to a blanketing effect of the wind.
 
\section {The hard component after the outburst}

   \begin{table}
   \caption[tabaccr2]{Assumed distances, derived X-ray
luminosities, optical magnitude V, ratio of optical to X-ray
luminosity, presence of long term
(l.t.) and orbital (orb.) variability for accreting novae
detected with ROSAT.
For V841 Oph and V446 Her we assumed a distance of 1 kpc .
 ``n'' means that  variability
is ruled out in the ROSAT exposures.
The magnitude in boldface for T Sco is in the B band. }
   \label{tabaccr2}
 \[
 \begin{array}{l c c c c c l}
 \hline
 $Nova$ & $ d$  & {\rm L}_x \times 10^{32} & $V$ &
 {\rm L_x/L_{opt}}  & {\rm l.t.  } & {\rm  orb.} \\
        & ${\rm (kpc)}$ & $erg  s$ ^{-1} & &
 \times 10^{-2} & & \\
   \noalign{\smallskip}
            \hline        \noalign{\smallskip}
 $V603 Aql$ & 0.43 &  2  & 11.7 & 1.4 & $x$ & $n$ \\
 $GK Per$   & 0.6  & 3.5  & 13.0 & 2.8 & $x$ &   \\
 $CP Pup$   & 0.7  & 0.59-5.9   & 15.0 & 4.8 & & $?$ \\
 $V841 Oph$ & 1.0 &   0.3  & 13.4 & 1.6 & $?$ & \\
 $CP Lac$    & 1.15 &  3.8  & 16.6 & 1.1 & $?$ & \\
 $T Sco$ & 9.1  &  33  &  {\bf 22.2} & 4.8 & & \\
 $DQ Her$    & 0.33 &  0.03  & 14.5 & 0.005 &  & \\
 $RS Oph$    & 1.2  &  0.28-1.64  & 12.5 & 0.2  & $x$  & \\
 $RR Pic$    & 0.26 & 0.02   & 12.2 & 0.06  & $x$  & $n$  \\
 $V446 Her$  & 1    & 0.72   & 18.0 & 3.3 & & \\
 $QU Vul$    & 2.6  & 0.18   & 19.0  & 1.7 & & \\
            \hline
         \end{array}
      \]
   \end{table}
   \begin{table}
   \caption[tabaccr3]{ Comparison of mass accretion rate
approximately  derived with
different methods (in M$_\odot$ $ \times 10^{-10}$ yr$^{-1}$).
 References are in the
text.  The second column indicates number of years elapsed between the
outburst and the X-ray observation.}
   \label{tabaccr3}
 \[
 \begin{array}{l c c c c  l}
 \hline
            \noalign{\smallskip}
 $Nova$ &  $years$ & {\rm \dot M_x(10) } &  {\rm \dot M_{IR}(10)}        & {\rm
\dot M_{opt}(10)}  & {\rm \dot M_{UV}(10)} \\
 \noalign{\smallskip}
            \hline
 $V603 Aql$ & $73-74$ &  1       & 1 & 10  & 100  \\
 $GK Per$  & 92 &  1.6   & & 100 &  \\
 $CP Pup$  & 50 & 0.35-3.5  & & 0.3 &   \\
 $V841 Oph$ & 152 & 1.8  & 30  & 1.7 & \\
 $CP Lac$ & 57 & 2.3  & 4000  & &  \\
 $T Sco$  & 132 &  20 & & &  \\
 $DQ Her$  & 58 & 0.02 & 6  & 300  & 50   \\
 $RS Oph$  & 7 & 0.1-1  & & & \\
 $RR Pic$ & 68 & 0.01  & 1  & & \\
 $V446 Her$  & $32-33$ & 0.04   &      4  & & \\
 $PW Vul$   & $7-9$ & < 6  & 34  & &  \\
 $V1500 Cyg$ & $26-27$ & <  0.04  & & 0.03 & 0.035 \\
 $QU Vul$ & 5 & 0.01   & & & \\
            \hline
         \end{array}
      \]
   \end{table}

In this and in the next section we will  use the adjective ``hard''
for the ROSAT range 0.8-2.4 keV.
Almost all the novae in outburst during the ROSAT
lifetime were observed.  Four out of the seven Galactic novae observed
with the PSPC within two years from the outburst
showed hard X-ray emission: V838 Her (N Her 1991)
 (see Lloyd et al. 1992), V351 Pup (N Pup 1991),
V3890 Sgr, and initially also  V1974 Cyg.
They represent 60\% of the available targets. 
Six other novae (V382 Vel, PW Vul,
QV Vul, RS Oph, LMC 1992 and 1995),  were observed to be bright in
X-rays either
with the ROSAT HRI or with other satellites within
two years from the outburst, although only V382 Vel was observed with
spectral resolution. To these we have to add at least
one LMC nova, LMC 1992, detected
(see Table 2) in the fourth month post-outburst (even if
undetected in the following month). Unfortunately,
for the hard X-ray emission the
observations in the LMC cannot add a large amount of
 information. At a distance d$\simeq$50 kpc,       
 sources with luminosities L$_{\rm x} \leq 10^{34}$ erg s$^{-1}$
(typical for the novae observed as hard sources with ROSAT)
are detected with exposures longer than 10$^4$ s, which were rarely
done.

    Looking more in detail at the observation of the individual objects,
 it is likely that ROSAT did not cover the whole 
 X-ray spectrum of the shocked nebula in the first few months post-outburst.
 This appears to be quite evident for V382 Her,
observed only after 6 days, and for the observations of  V1974 Cyg
 in the first 5 months. Fits of  thermal
models (available in EXSAS) to the observed
spectra (6 days post-outburst for V383 Her, and 63 days for
V1974 Cyg),
 leave undetermined whether the plasma temperature was in the
5-15 keV range with N(H)$\leq$10$^{22}$ cm$^{-2}$,
or instead kT$\simeq$1-2 keV  with N(H)$>$10$^{22}$ cm$^{-2}$. We consider the
first possibility more likely, at the light of the V382 Vel
BeppoSAX and ASCA observations (Orio et al. 2001, Mukai \& Ishida 2001).
Of course, high N(H) cannot be ruled
out: in V382 Vel it  was due to internal absorption of the
ejecta, rich in C and O. Probably
the X-ray emission came initially from
deeply inside the ejecta, or from  a very clumpy material.
 V1974 Cyg was observed again before the super-soft           
X-ray emission became dominant. At day 91
the plasma temperature was not well constrained yet, and N(H)$\leq$3.9
$\times$ 10$^{21}$ cm$^{-2}$,  but in one observation of day 147,
the
plasma temperature was bound to be below 1.4 keV (3$\sigma$ upper limits).
 The temperature was quickly cooling {\it or} the
shell absorption was rapidly decreasing, {\it or} both things were happening.
Also for V382 Her, Orio et al. (2001) found kT$\leq$1 keV half a year 
post-outburst.  As the temperature of the emitting plasma 
was decreasing, the total
hard flux increased in the case of N Cyg 1992  in the
 first few months
after the outburst (see also Balman et al., 1998).
 This could be due either to the thinning
of the ejecta to X-rays, or to an increasingly large shocked region.

How long does it take for the shocked nebula to cool? V3890 Sgr,
 the only recurrent nova in outburst during the
operation period of ROSAT, has a red giant companion and
other characteristics in common with RS Oph, the first
hard X-ray source observed among post-novae. This nova
was detected in the ROSAT All-Sky-Survey
(Orio et al. 1992a) but it  did not appear as        
 an X-ray source in  serendipitous pointings after 11 and 17 months,
so the emitting shell must have cooled by then.
We also know that  the cooling time
 can definitely be longer than 11 months.
 V351 Pup (N Pup 1991) was X-ray bright 16 months after the outburst
(see also Orio et al. 1996), but 
the X-ray source had faded in a later observation
done with ASCA in 1998 (Orio \& Mukai, 2001). At 16 months the best fits
of a thermal plasma were consistent with lack of intrinsic 
absorption, kT$\simeq$1 keV, and X-ray luminosity 
of a few times 10$^{33}$ erg s$^{-1}$. In exceptional
 cases, we find hard X-ray emission from the shell even
tens of years after the outburst.
For RR Tel, after 43 years the hard X-ray
component is ascribed to colliding winds
producing a shock (Jordan et al. 1994, Contini \& Formiggini 1999).
We fit the data with flux of a few 10$^{-14}$ erg cm$^{-2}$ s$^{-1}$
and plasma temperatures 200 and 350 eV
in the two observations respectively.
 GK Per had a large amount of
circumstellar material, most likely residual of a planetary
nebula phase.  In the best fit with
two components, we find that the total flux of the softest component is
$\simeq$25\%  the total flux (approximately 8 $\times 10^{-12}$
erg cm$^{-2}$ s$^{-1}$), with kT$\simeq$ 200 eV
and N(H)$\simeq$ 10$^{21}$ cm$^{-2}$ (consistent with the
 measured value E(B-V)$\simeq$0.1).
Balman \& \"Ogelman (1999) indeed found in a later,
spatially resolved HRI observation, that
a fraction 20-25\% of the total flux is due to the
extended shell and this component is the softest one.                           
\section{The accretion process: signatures in X-rays}

The  novae included in Table 4, except RR Tel and partially
GK Per, were observed in X-rays at quiescence because of {\it accretion}.
However, only 11 out of 81 quiescent Galactic novae observed 
in the pointed observations were detected. Only the
detection of CP Lac is entirely new.   
Comparing catalogs
of classical novae (D\"urbeck 1988) and of dwarf novae
(Khruzina \& Shugarov, 1991) we find that classical novae are on average only 
$\simeq$ 2 magnitude fainter at quiescence, despite  a larger
average distance by an order of magnitude. The X-ray luminosity 
 of novae however, at least in the ROSAT range, 
is not higher than for dwarf novae.  For the detected objects 
L$_{\rm x} = 10 ^{30-33}$ erg s$^{-1}$ (see Table 5).
We notice that the novae that appear as X-ray luminous 
at quiescence were all {\it fast} novae in outburst.
We see in Fig. 1 a complex and large  variety of spectral
characteristics, even in the relatively narrow ROSAT range. The spectra 
vary from very soft, like RR Pic, to
to hard, like CP Pup. Also the  ratio of L$_{\rm x}$/L$_{\rm opt}$
varies, as shown in Table 5.

  For the non detected Galactic  novae,
assuming a plasma temperature kT$\simeq$ 1 keV for the accretion disk, the
3$\sigma$ upper limits on the ROSAT X-ray flux are always below
a few 10$^{-13}$ erg cm$^{-2}$ s$^{-1}$.
At an average distance 1-2 kpc this corresponds to 
L$_{\rm x} \leq 10^{30}$ erg s$^{-1}$. 
Since most IP are at higher X-ray luminosity
and exhibit a  hard X-ray flux component,  which is not 
significantly affected by interstellar absorption,
this is a general indication that  IP do not seem to be very frequent among
novae (we caution however that DQ Her, the 
prototype of intermediate polars, has {\it lower} X-ray luminosity
 than the rest: are perhaps some
 IP-novae unusually under-luminous in X-rays?).

The All-Sky Survey detection of a source in the spatial error box 
of both  V382 Sco and  V720
Sco (Orio et al. 1992a) is ruled out examining  the pointed observations
(or the source is variable at quiescence).
Only marginal detections were
obtained for V446 Her and V841 Oph (observed
only during the All-Sky Survey) and in the pointed observations
of DQ Her,  T Boo, and CP Lac.

For RR Pic the super-soft component is fitted 
with the  a blackbody at T$_{\rm BB} \simeq$ 25 eV
 and F$_{\rm bol} \simeq 2 \times 10^{-13}$ erg cm$^{-2}$ s$^{-1}$,
 much too low to be originated on the WD surface. 
 There is an additional, less luminous  hard component. The parameters are
 not well constrained, although a Raymond-Smith model 
with kT$\simeq$ 1 keV and flux F$_x \simeq 5 \times 10^{-14}$
erg cm$^{-2}$ s$^{-1}$ fits the data reasonably. We favour the
interpretation of van Teeseling et al. (1996)
that the system is a polar, despite the low X-ray luminosity.
On the basis of the
X-ray spectral characteristics, this is the only new suspect
magnetic system  in all our sample.
For  GK Per, already known instead  to be an IP, 
a reasonable fit to the harder spectral component
(the one attributed to accretion, and corresponding to 
$\simeq$75\% of the total X-ray flux)
gives N(H)=2 $\times$ 10$^{22}$ cm$^{-2}$,
kT=870 eV and flux 6.5 $\times  10^{-12}$ erg cm$^{-2}$ s$^{-1}$.
N(H) is thought to be partially due to the nebula.

 We are puzzled by the different behaviour of
the two X-ray brightest novae,  V603 Aql and CP Pup.
They displayed, however,
similar optical characteristics in outburst (Payne-Gaposhkin 1957)
and at quiescence (the super-hump phenomenon).
We rule out significant changes in the X-ray luminosity and spectrum of CP Pup
from {\it Einstein} times.
The opposite is true for  V603 Aql.
In the early 80's the count rate measured with Einstein
varied by more than a factor of 3 within a couple of years (see Table 4). 
Drechsel et al. (1987)  estimated
L$_{\rm x} \simeq$3 $\times$ 10$^{33}$ erg s$^{-1}$ and 
plasma temperature kT=20-30 keV.
There must be, however, even 
three components in the X-ray flux of this nova, because
Mukai (2000) showed that the ASCA spectrum can be fitted
with two components at 1 and 9 keV. ROSAT is sensitive only to the lowest
energy component, which instead must have been missing for CP Pup.
\begin{figure*}
\centering
\includegraphics[angle=-90,width=8cm,clip]{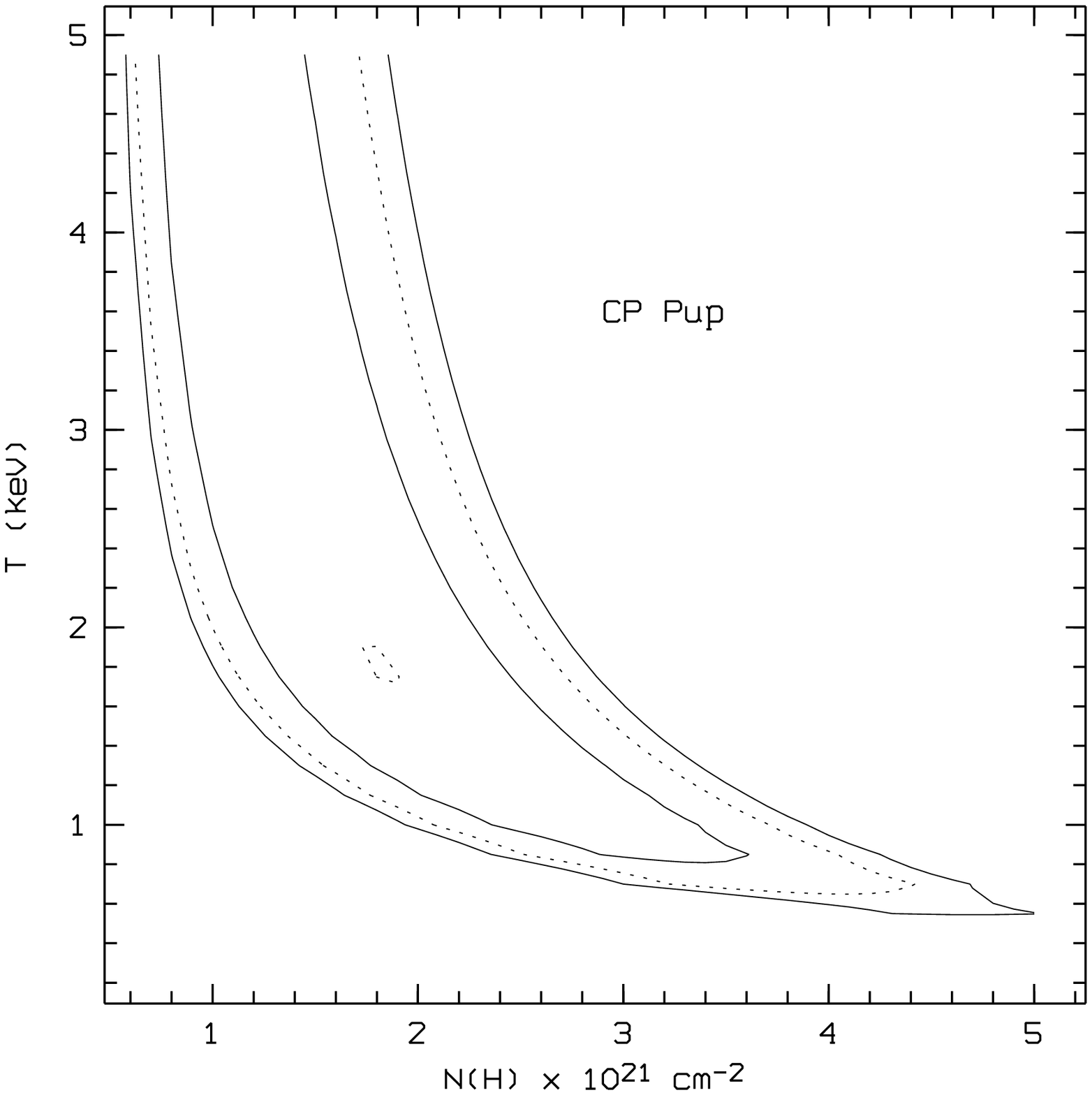}
\quad \includegraphics[angle=-90,width=8cm,clip]{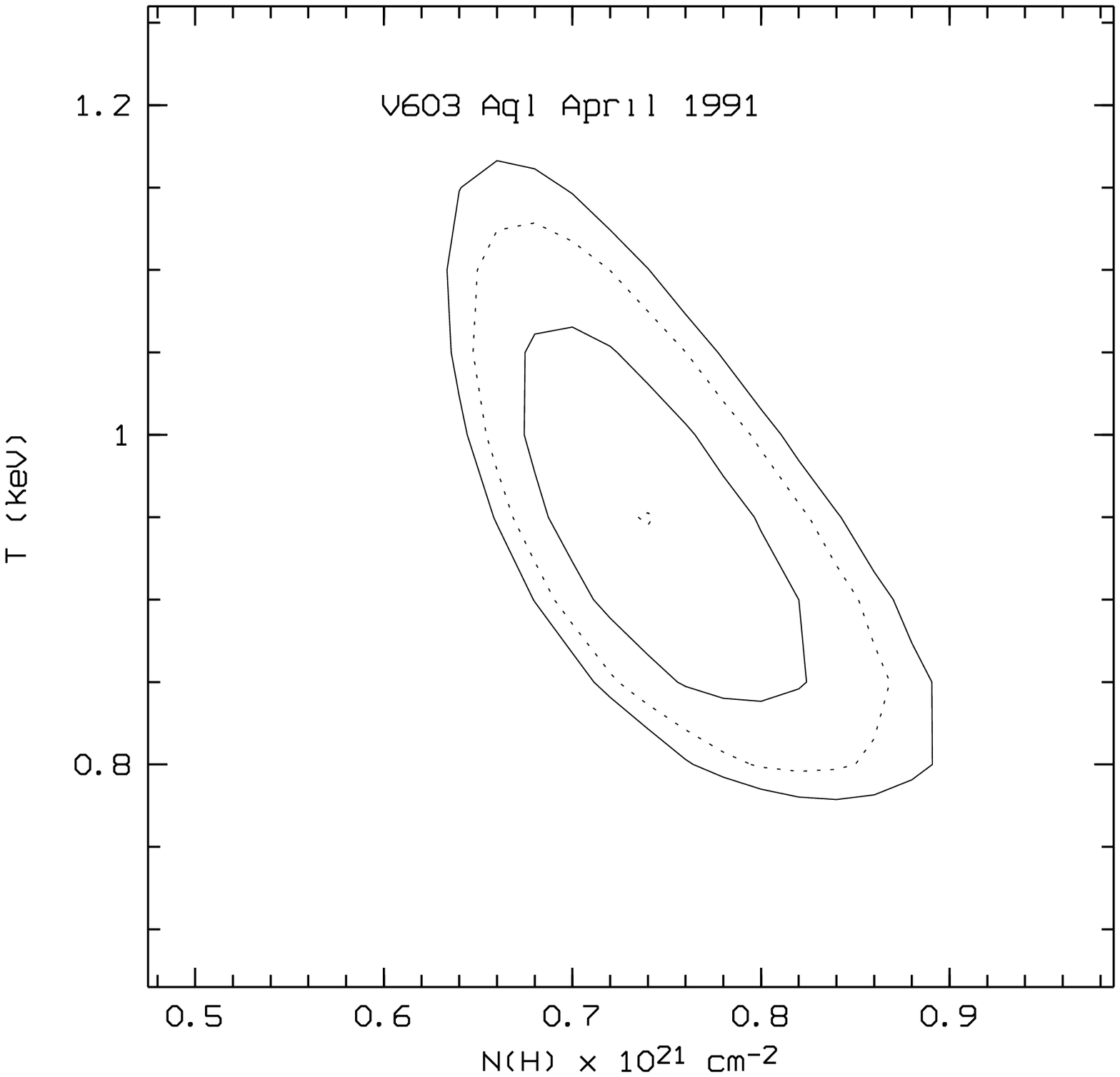}

\medskip
\includegraphics[angle=-90,width=8cm,clip]{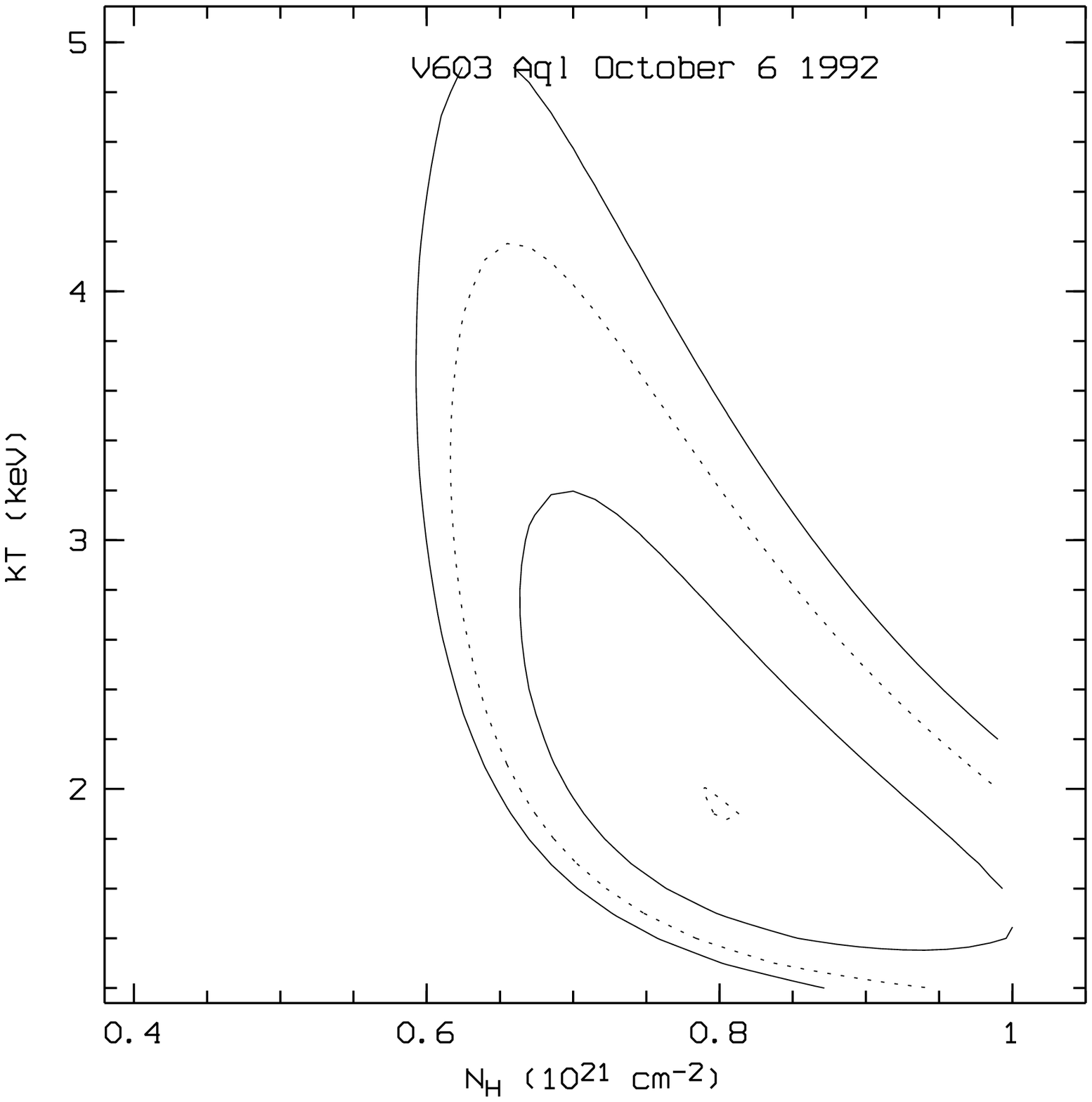}
\quad \includegraphics[angle=-90,width=8cm,clip]{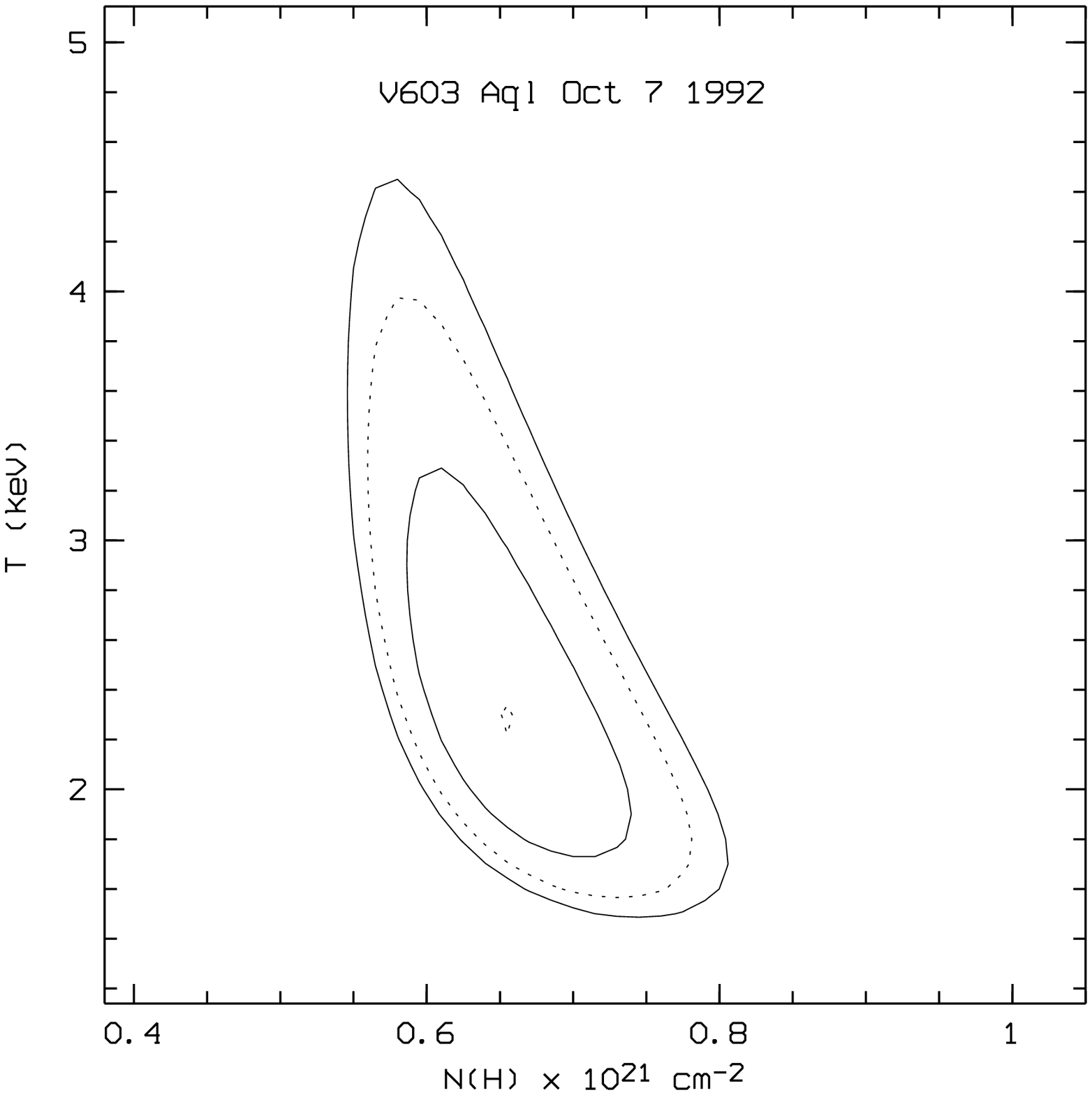}
\caption[accr]{ 1,2 and 3$\sigma$  confidence
 contours in the N(H) versus
 T (plasma temperature) plane for CP Pup and for V603 Aql in
three different PSPC observations: in April
1991
and in two of the short observations taken in 1992, done on October 6 and 7
respectively (see Table 3).
The small dotted contours indicate the best fit values. The best  fit flux
is 2.1 $\times$ 10$^{-12}$ erg cm$^{-2}$ s$^{-1}$ for
CP Pup, and for the three
V603 Aql observations shown we obtained
 7.9 $\times$ 10$^{-12}$ erg cm$^{-2}$ s$^{-1}$,
         10$^{-11}$ erg cm$^{-2}$ s$^{-1}$,
and 2.8 $\times 10^{-11}$ erg cm$^{-2}$ s$^{-1}$ respectively.}
\label{accr}
\end{figure*}
In Fig. 2 we show the spectral probability contours
in the N(H) vs. kT plane for the only
pointed observation of CP Pup, and for three representative
different observations of V603 Aql.
From 1991 to 1992 the PSPC
count rate measured for V603 Aql increased by a factor of 3,
and during the 1992 observations it varied by a factor of 60\%
 within 5 days.  The spectrum in the PSPC range
 hardened in 1992 but did not seem to
vary very much over hours and days (see Fig. 2).  
 Drechsel et al. (1987) found irregular small scale
variations 
over-imposed on the long scale one, and  eclipses along the
orbital period. In the light curve  derived from 27 successive
 observations
done in 1992 for approximately 1000 s each and spaced few hours apart
  (Fig. 3), no variations with the
 orbital periods (spectrometric and photometric) in the ROSAT range
are present at a 3$\sigma$ confidence level. There
is still flickering on short ($\simeq$10-100 s) time scales. 
The ROSAT-PSPC light curve  over 120 hours of observations in 1992 
(shown in Fig. 3) gives a very interesting indication of a 
$\simeq$60 hour periodicity, which needs to be verified but is 
observed at optical wavelengths and corresponds to the beat
 period (Udalski \& Schwarzenberg-Czerny, 1989).
For CP Pup (see also Balman et al. 1995),
as the open contours seem to imply, there is no soft component 
and most likely
the plasma temperature is above the ROSAT range.
X-ray modulations with the orbital period (Balman et al. 1995) 
and the high X-ray luminosity  seem to be 
the signatures of an IP, however this interpretation
is ruled out by Patterson \& Warner (1998).
\begin{figure}
\includegraphics[width=8.7cm]{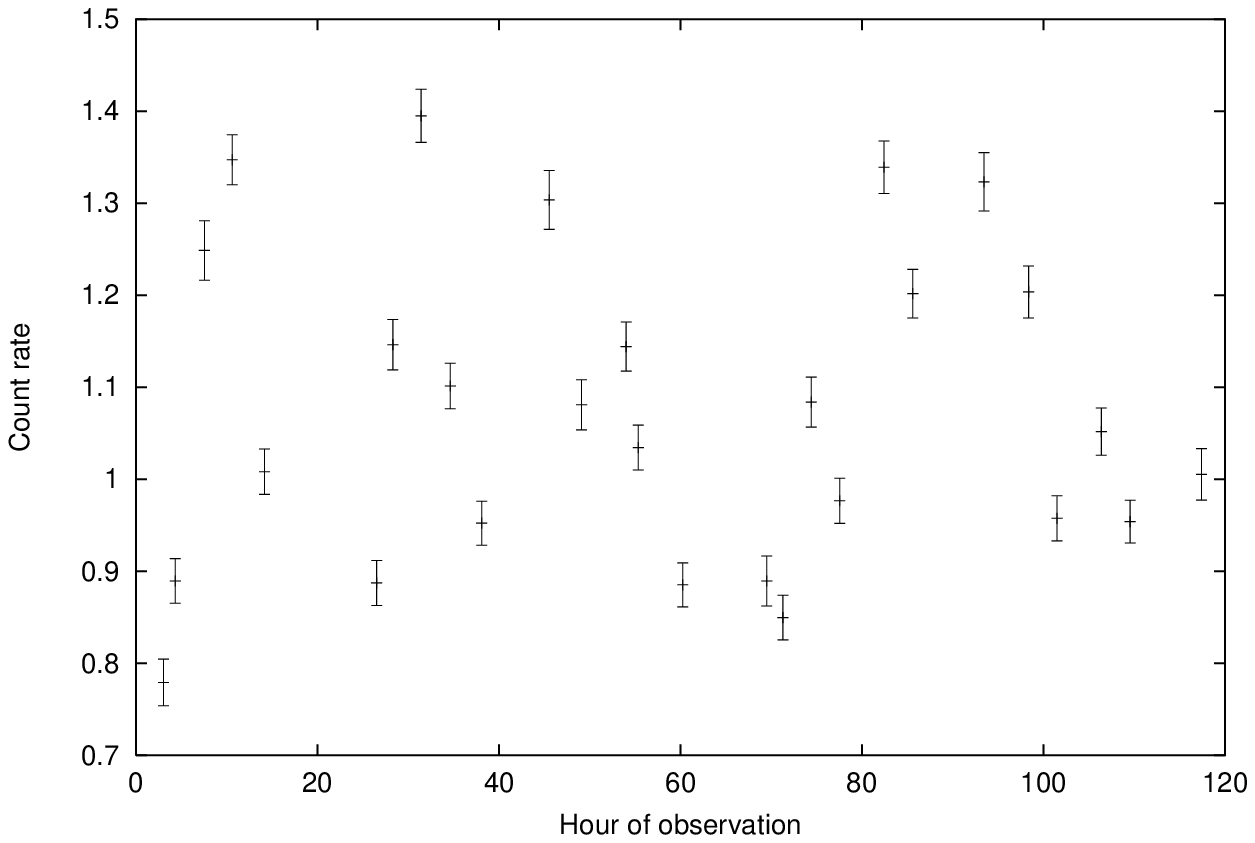}
\caption[lc]{The ROSAT PSPC light curve of V603 Aql
observed intermittently for almost 120 hours in October 1992. Each point
 is the average of an observation lasting for about 1000s.}
\label{lc}
 \end{figure}
 
Altogether, not only V603 Aql but
5 out of 11 objects seem to be variable X-ray sources: 
first of all RS Oph, which was almost three times less
luminous in an unpublished serendipitous observation in 1991 than when
observed by us 1992 (Orio 1993). GK Per, V841 Oph and RR Pic also 
show hints of long term variability. 
The IPC-PSPC count rate conversion is done with PIMMS
assuming kT=1 keV. 

The most important conclusion
to be drawn from the ROSAT database is
that there is no significant difference in the X-ray luminosity
of dwarf novae and novae. Otherwise, 
the selection effect due to the distance would be compensated 
by the larger X-ray luminosity even at quiescence. 
As Table 5 shows
there is also no clear correlation between X-ray luminosity L$_{\rm x}$ and
the visual magnitude V at minimum. The ratio L$_{\rm x}$/L$_{\rm opt}$, where
L$_{\rm x}$ is the X-ray luminosity in the 0.2-2.4 keV range, varies
greatly in the detected object and is low:
 L$_{\rm x}$/L$_{\rm opt}<$0.03 for most detected novae, and
L$_{\rm x}$/L$_{\rm opt}<$0.01 for the non-detected objects. 
 For those CV which, like most novae show strong emission
lines at quiescence, and are likely to have optically
thin disks, Patterson \& Raymond 1985) find typically
L$_{\rm x}$/L$_{\rm opt} \simeq$0.03. This ratio is only $\leq$0.01
 in  erupting dwarf novae, where 
most of the boundary layer emission is thought to be radiated
in the EUV range (Mauche et al. 
1998).  Is this also true for classical novae? 

At least, as we show in Table 6, we rule out that  
the mass transfer rates inferred from the X-ray luminosity in 
the ROSAT range are representative. 
The accretion rates $\dot {\rm M}_{\rm x}$ in the table were 
 calculated assuming a  variation of the analytical formula
of Patterson \& Raymond (1985) adapted to the ROSAT range:
$$\dot {\rm M}_{\rm x} = {{\rm L_x}(0.2-2.4 {\rm ~keV})
 \over 11.5 \times 10^{31} {\rm M}^{1.8}_{0.7}}
{\rm ~erg  ~s}^{-1}$$
 where M$_{0.7}$ = M$_{\rm WD}$/0.7 M$_\odot$.
In this formula a Raymond-Smith model of thermal plasma with
kT=2 keV and N(H)=10$^{21}$ cm$^{-2}$ is assumed, and
the fraction of gravitational energy which is re-radiated
by the boundary layer 
is 0.5.   The formula works for other CV for 
$\dot {\rm M} \leq 10^{-16}$ g s$^{-1}$.
Assuming for the sake of simplicity always 
 M$_{\rm WD}$=1.3 M$_\odot$, we derived the approximate mass transfer
 rates $\dot {\rm M}_{\rm x}$ in Table 6.  $\dot {\rm M}$ 
 inferred from infrared observations (Weight et al. 1994),
from the optical magnitude and its the rate
of decline (D\"urbeck 1992), from the optical magnitude,
and from UV observations
(Patterson 1984 and references therein)
often differs from $\dot {\rm M}_{\rm x}$ by 
much more than one order of magnitude (which would still 
be justified by a lower WD mass or lower fraction
of re-radiated energy). 
Whether this ``missing boundary layer emission'' 
is due to the accretion energy being re-radiated mostly
{\it below} 0.2 keV, is a possibility that we cannot rule out. 

There is no correlation of the derived $\dot {\rm M}_{\rm x}$ with the
 post-outburst
age as predicted by the hibernation theory (Shara et al. 1986).
  According to this theory, accretion occurs at a very high rate in the last
50-100 years before the outburst and for at least 
$\simeq$10 years after.  Concerning the time {\it before} the outburst,
serendipitous observations were done with the ROSAT PSPC
in the last 1 to 4 years before the outburst for only three Galactic novae,
  V1419 Aql, BY Cir and  CP Cru. The 3$\sigma$ flux
upper limits are  of
the order of 10$^{-13}$ erg cm$^{-2}$ s$^{-1}$. 
We derive L$_{\rm x} \leq 2 \times 10 ^{30}$ erg s$^{-1}$ 
for  CP Cru and BY Cir (arbitrarily assuming  
a d=1 kpc), and L$_x \simeq 10^{31}$ erg s$^{-1}$ for N Aql 1993 
(assuming d=2.7 Kpc, from Kamath et al. 1997). 
Two of the novae were {\it moderately fast} and one was
{\it fast} (according to the theory, the faster the nova
in outburst the larger the accretion rate). 

\section{Conclusions}

1.
Only $<$20\% of classical
and recurrent novae have been observed as super-soft X-ray sources
at some snapshot in time. 
If turn-off of the central source occurs within a few months, the 
accreted envelope is depleted more efficiently than 
foreseen by the models. We suggest three possible causes 
to explain this fact:
a) a line driven wind depleting hydrogen burning material
 at the end of the outburst (Starrfield, 1997,
private communication)
b) a magnetically driven wind (e.g. Orio et al. 1992b), c) a wind 
due to the secondary  in the ejected shell
(i.e. ``common envelope'' type of phenomenon).\\

\noindent 2. More than half of all classical and recurrent novae were
observed as transient hard X-ray sources 
in the first two years after the outburst. The origin is
attributed to shocked ejecta. The shell
may cool within a couple of years but it may also 
emit X-rays for up to a century under particular conditions,
like previously existing circumstellar material (GK Per) or a 
wind continuing after the outburst (RR Tel). \\

\noindent 3. The average X-ray luminosity of
quiescent novae in the ROSAT range
is not higher than L$_{\rm x} \simeq$ few times 10$^{30}$ erg s$^{-1}$,
measured on average for dwarf novae by Verbunt et al. (1997).
Our conclusion is based mainly on upper limits, because
the majority of the Galactic novae were not detected. We remark that 
the X-ray brightest classical novae detected in quiescence are all
{\it fast novae}.\\ 

\noindent 4. There is no clear correlation  of X-ray and visual luminosity.
For the majority of quiescent novae 
L$_{\rm x}$/L$_{\rm opt} \leq$0.01. 
The mass transfer rates derived from the X-ray luminosity are much
 lower than those inferred from other wavelength ranges,
 implying a ``missing boundary layer''.
 Whether this is due simply to the fact that the accretion energy is not
 re-radiated in X-rays but almost
entirely in the EUV range, like for DN in outburst
(Mauche 1998), at the moment cannot be proved or disproved.\\

\noindent 5. There
 is no correlation between X-ray luminosity and time elapsed 
before or after the outburst.\\ 

\noindent 6. The nova that was best studied with the ROSAT PSPC at
quiescence, V603 Aql (also the most luminous)  has variable X-ray
spectrum and luminosity on time scales of 1 year and 
of few days. 
The recurrent nova RS Oph and a few classical novae 
also seem to have been at different
X-ray luminosity (or spectral energy distribution)
 during the ROSAT lifetime.
Most likely the luminosity level, or the spectral distribution,
was also different in the observations done with {\it Einstein}
approximately 10 years earlier.
A self-consistent theory of accretion in classical novae has to
take this  X-ray variability into account. 
Monitoring the long term X-ray light curve of quiescent novae
 in an efficient way
and explaining its characteristics is a challenge we have ahead. 
 
\begin{acknowledgements}
 Both M.O and H.\"O acknowledge support of the Graduate School of  
the University of Wisconsin during the course of this project. 
The project was also partially supported by the Italian Space Agency ASI and
by the NASA grant 98-03-ADP-042. We thank Jim MacDonald for the
white dwarf atmospheric models.
\end{acknowledgements}

\end{document}